\pdfoutput=1
\documentclass[journal]{IEEEtran}
\usepackage[T1]{fontenc} 
\usepackage{amsmath} 
\usepackage{amssymb} 
\usepackage[pdftex]{graphicx} 
\usepackage{color}

\newcommand{\diag}{\mbox{\rm diag}}
\newcommand{\ssst}{\scriptscriptstyle}
\newcommand{\gam}[1]{\mbox{$\gamma_{\!\ssst#1}$}} 

\newcommand{\rem}{\paragraph*{Remarks}} 

\newcommand{\beq}[1]{\begin{equation}\label{#1}}
\newcommand{\eeq}{\end{equation}}
\newcommand{\bfB}{\mbox{\bf B}}
\newcommand{\bfD}{\mbox{\bf D}}
\newcommand{\bfE}{\mbox{\bf E}}
\newcommand{\bfH}{\mbox{\bf H}}
\newcommand{\bfS}{\mbox{\bf S}}

\begin{document}

\title{Gain Scaling for Multirate Filter Banks}

\author{%
	\thanks{Los Alamos National Laboratory is operated for the  U.S.\ Department of  Energy by Los Alamos National Security LLC  under Contract No.\ DE-AC52-06NA25396. This work was supported in part by the Los Alamos Laboratory-Directed Research and Development Program.}%
	\IEEEauthorblockN{Christopher M.\ Brislawn}%
	
	\IEEEauthorblockA{%
		Computer, Computational \& Statistical Sciences Division\\%
		Los Alamos National Laboratory\\%
		Los Alamos, NM 87545--1663 USA\\%
		E-mail: {\tt brislawn@lanl.gov}%
	}
}

\IEEEspecialpapernotice{(Invited Paper)}

\maketitle
\thispagestyle{empty} 
\begin{abstract}
Eliminating two trivial degrees of freedom corresponding to the lowpass DC response and the highpass Nyquist response in a two-channel multirate filter bank seems simple enough.  Nonetheless, the ISO/IEC JPEG 2000 image coding standard manages to make this mundane task look totally mysterious.  We reveal the true meaning behind JPEG 2000's arcane specifications for filter bank normalization and point out how the seemingly trivial matter of gain scaling leads to highly nontrivial issues concerning uniqueness of lifting factorizations.
\end{abstract}

\begin{figure}[h]
  \begin{center}
    \includegraphics{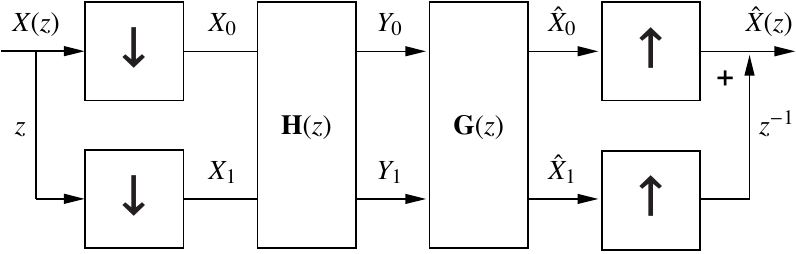}
    \caption{The polyphase-with-advance filter bank representation.}
    \label{DS_poly}
  \end{center}
\end{figure}
\section{Introduction}
\label{Intro}
The  starting point for studying lifting factorizations is the {\em polyphase-with-advance representation\/} for two-channel FIR multirate filter banks depicted  in Fig.~\ref{DS_poly}~\cite{DaubSwel98,BrisWohl06}.  
A {\em lifting factorization\/}~\cite{Sweldens96,DaubSwel98} is a cascade  decomposition of the analysis and synthesis polyphase matrices, $\mathbf{H}(z)$ and $\mathbf{G}(z)$, into alternating upper and lower triangular matrices with 1's on the diagonal and an FIR \emph{lifting filter} in the off-diagonal position.  

\begin{figure}[t]
  \begin{center}
    \includegraphics{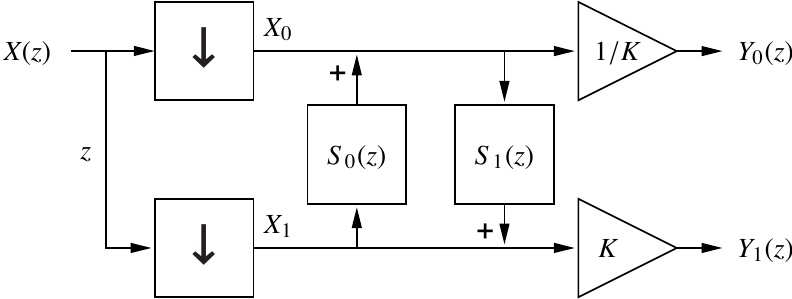}
    \caption{Analysis filter bank lifting decomposition with two lifting steps.}
    \label{irr_lift}
  \end{center}
\end{figure}
For instance, Fig.~\ref{irr_lift} depicts an analysis filter bank represented as a cascade of two ladder steps that alternately lift, or update, the demultiplexed input channels.  
The filters $S_0(z)$ and $S_1(z)$ are the lifting filters, and the   {\em scaling factor}, $K$,   specifies a relative gain normalization of the  channels.  Algebraically, this cascade corresponds to a factorization of the analysis polyphase matrix into 
lifting matrices:
\[
\bfH(z) = \mathbf{D}_K\,\bfS_1(z)\,\bfS_0(z)\,,
\]
where 
\[\mathbf{D}_K\equiv \diag(1/K,\,K)\]
 and
\[
\bfS_0(z) = \left[ \begin{array}{cc}
                1  & S_0(z) \\
                0  & 1
            \end{array}\right]\quad,\quad
\bfS_1(z) = \left[ \begin{array}{cc}
                1  & 0 \\
                S_1(z)  & 1
            \end{array}\right]\,.
\]
Daubechies and Sweldens~\cite{DaubSwel98} showed that \emph{any} two-channel FIR perfect reconstruction filter bank can be factored into such a form,
\begin{equation}\label{lifting_factorization}
\bfH(z) = \mathbf{D}_K\,\bfS_{N-1}(z)\cdots\bfS_1(z)\,\bfS_0(z)\,,
\end{equation}
where the matrices $\mathbf{S}_i(z)$ are alternating lower and upper triangular lifting matrices.

The {\em update characteristic}, $m_0$, of the first lifting step in Fig.~\ref{irr_lift} is  ``lowpass'' ($m_0=0$), while the update characteristic, $m_1$, of the next lifting step is ``highpass'' ($m_1=1$).  Since the update characteristic of successive  steps alternates, it suffices to transmit  the update characteristic, $m_{\it init}$, of the initial synthesis lifting step, which is the same as the update characteristic of the {\em last\/} analysis lifting step.

\begin{figure}[b]
  \begin{center}
    \includegraphics{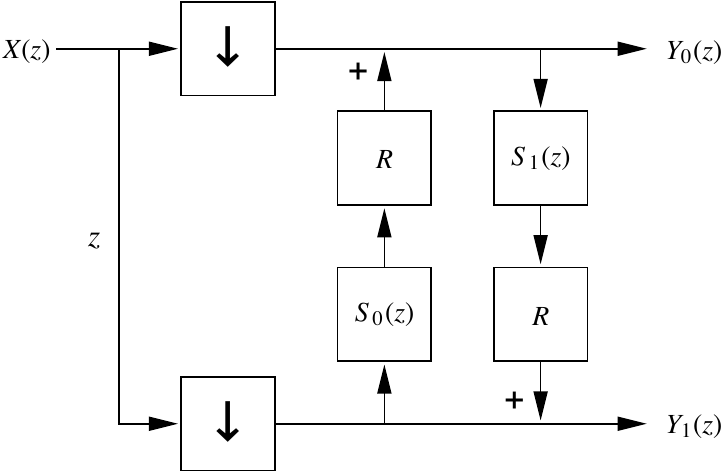}
    \caption{Reversible analysis filter bank lifting decomposition.}
    \label{rev_lift}
  \end{center}
\end{figure}
If the impulse response taps in the lifting filters are dyadic rationals, the lifting updates can be applied to an integer-valued input in integer arithmetic by rounding the updates with a rounding function, $R$, immediately before applying them, as shown in Fig.~\ref{rev_lift}.  Such implementations are called \emph{reversible} and yield bit-perfect inversion in fixed precision arithmetic.  Filter banks that are not reversible are called \emph{irreversible}.  The gain scalings $K$ and $1/K$ shown in Fig.~\ref{irr_lift} are not allowed in reversible implementations because they cannot be inverted losslessly in fixed precision arithmetic.

\section{Filter Banks in JPEG~2000}
\label{JPEG2000}
Filter banks are specified in the ISO/IEC JPEG~2000 image coding standard (International Standard~15444-x) as a sequence of FIR lifting filters, signaled via their impulse responses $s_i(n)$, and a scaling factor $K$.
JPEG~2000 Part~1~\cite{ISO_15444_1,TaubMarc02} provides just two predefined filter banks, one reversible and one irreversible, but users may specify arbitrary two-channel FIR filter banks in JPEG~2000 Part~2~\cite{ISO_15444_2} (henceforth just ``Part~2'') via user-defined lifting filters.  To simplify quantization and entropy coding, however,  the standard specifies normalization requirements that user-defined JPEG~2000 filter banks must satisfy.
We now outline the mathematical basis of these normalization requirements, as detailed in~\cite{BrisWohl07}.

\subsection{JPEG~2000 Filter Bank Normalization Requirements}
\label{JPEG2000:Requirements}
The normalization requirements in JPEG~2000 Part~2 Annexes~G and~H are given in the following form.  Let 
\[ D_i=\sum_n s_i(n)=S_i(1) \]
be the DC response of the $i^{th}$ lifting filter, $S_i(z)$.  Define a two-step recursion, 
\begin{equation}\label{recursion}
B_i = D_i B_{i-1}+B_{i-2}\quad\mbox{for  $i=0,\ldots,N-1$,} 
\end{equation}
with initial conditions 
\[B_{-1}=B_{-2}=1\,.\]
  Then an irreversible user-defined  filter bank must satisfy
\begin{eqnarray}\label{irr_normalization}  
B_{N-2}	&=& K\quad\mbox{if $m_{\it init}=1$, otherwise}  \\
B_{N-1}	&=& K\quad\mbox{if $m_{\it init}=0$.}\nonumber
\end{eqnarray}
The requirement for reversible filter banks is
\begin{eqnarray}  \label{rev_normalization} 
B_{N-2}	&=& 1\quad\mbox{if $m_{\it init}=1$, otherwise}  \\
B_{N-1}	&=& 1\quad\mbox{if $m_{\it init}=0$.}\nonumber
\end{eqnarray}

These requirements, which total less than one page of content in Part~2~\cite{ISO_15444_2}, are not derived or explained in the standard and are extremely cryptic as written.  

A natural normalization to impose on filter banks would have been to fix the lowpass DC response and the highpass Nyquist response of the filters $H_0(z)$ and $H_1(z)$.  Since filter banks are defined via lifting~(\ref{lifting_factorization}), the DC and Nyquist responses are determined by the lifting filters and the scaling factor, $K$.   In the case of filter {banks} satisfying the perfect reconstruction condition,
\begin{equation}\label{DS_FIR_PR}
\det\bfH(z) = 1\,,
\end{equation}
one  degree of freedom  is consumed by normalizing the polyphase determinant.  The JPEG~2000 standard can therefore only normalize one more trivial degree of freedom, and the unstated goal of the JPEG~2000 normalization requirements given above is to achieve a lowpass DC response of unity:
\begin{equation}\label{lowpassDCconstraint}
H_0(1)=1\,.
\end{equation}
The highpass Nyquist response will then be determined by~(\ref{DS_FIR_PR}) and~(\ref{lowpassDCconstraint}), as explained  in~\cite{BrisWohl07}.

\subsection{Calculating the Lowpass DC  Response}
\label{JPEG2000:Responses}
The relationship between the analysis filters, $H_0(z)$ and $H_1(z)$, and the  polyphase analysis matrix, $\bfH(z)$, can be stated in  matrix-vector arithmetic.
Using an underscore for vectors, the relationship~\cite[Formula~(9)]{BrisWohl06} can be written
\beq{Hi_poly_matrix}
\underline{H}(z)\;\equiv\; \left[ \begin{array}{l}
        H_0(z)\\
        H_1(z)
    \end{array} \right]
   \; = \;
    \bfH(z^2)
    \left[ \begin{array}{l}
        1\\
        z
    \end{array} \right]\;.
\eeq

We will use this formula to calculate the DC responses of the filters $H_0(z)$ and $H_1(z)$ corresponding to a lifting factorization~(\ref{lifting_factorization}).  Let $\bfE(z)$ denote the {\em unnormalized\/} cascade of lifting steps (without $\mathbf{D}_K$),
\[ \bfE(z) \equiv \bfS_{\ssst N-1}(z)\cdots\bfS_1(z)\,\bfS_0(z)\;, \]
so that
\begin{equation}\label{unnormalized}
\mathbf{H}(z)=\mathbf{D}_K\,\mathbf{E}(z)\,.
\end{equation}
By~(\ref{unnormalized}) the lowpass filter is 
\[H_0(z)=(1/K)E_0(z)\]
so the lowpass DC condition~(\ref{lowpassDCconstraint}) becomes
\begin{equation}\label{lowpassDCconstraint2}
E_0(1)=K\,.
\end{equation}

Since $K$ is transmitted in the JPEG~2000 Part~2 codestream along with the user-defined lifting filters, this condition is actually a functional relationship between $K$ and the lifting filters that must be satisfied in order to produce a compliant Part~2 codestream.  To ensure that~(\ref{lowpassDCconstraint2}) is satisfied, we thus need to compute $E_0(1)$ in terms of the lifting filters.

Let $\bfE^{(n)}(z)$ denote the $n^{th}$ partial product:
\[ 
\bfE^{(n)}(z) = \bfS_{\ssst n}(z)\cdots\bfS_0(z)
\quad\mbox{for}\quad \mbox{$n=0,\ldots,N-1$.} 
\]
The vector of corresponding lowpass and highpass scalar filters is given by~(\ref{Hi_poly_matrix}):
\begin{eqnarray}
\left[ \begin{array}{l}
        E^{(n)}_0(z)\\
        E^{(n)}_1(z)
    \end{array} \right]  & \;\equiv\; & \underline{E}^{(n)}(z) 
    \;=\; 
    \bfS_{\ssst n}(z^2)\cdots\bfS_0(z^2)
    \left[ \begin{array}{l}
        1\\
        z
    \end{array} \right] \nonumber\\
    & = &
    \bfS_n(z^2)\,\underline{E}^{(n-1)}(z) \label{En_poly_vector}  \;.
\end{eqnarray}
Using~(\ref{En_poly_vector}), the DC response vector can be constructed via the recursion
\beq{DC_gain_vector_recursion}
\underline{E}^{(n)}(1) = \bfS_n(1)\,\underline{E}^{(n-1)}(1)\quad\mbox{for $n=0,\ldots,N-1$,}
\eeq
where
\[ 
\underline{E}^{(-1)}(1) \equiv 
    \left[ \begin{array}{l}
        1\\
        1
    \end{array} \right]\;.
\]

JPEG~2000 does not employ matrix-vector arithmetic, however, so we now need to put the vector recursion~(\ref{DC_gain_vector_recursion}) into the form of a scalar recursion.  The matrix $\bfS_i(1)$ contains the value $D_i=S_i(1)$, and it is shown in~\cite{BrisWohl07} that if one defines  $B_n$ to be the {\em most recently modified entry\/} in the DC response vector, $\underline{E}^{(n)}(1)$, then~(\ref{DC_gain_vector_recursion}) can be reduced to the scalar recursion~(\ref{recursion}) appearing in the Part~2 normalization requirements.

How do we translate the lowpass DC gain condition~(\ref{lowpassDCconstraint2}) into the Part~2 requirement~(\ref{irr_normalization})?  As shown in~\cite{BrisWohl07}, the requirement~(\ref{irr_normalization}) is determined by the parameter $B_n$  that gives the value of the final lowpass DC response, $E_0(1)$.  This in turn can be stated  in terms of the update characteristic, $m_{\it init}$, of the last analysis lifting step, as is done in~(\ref{irr_normalization}).  Finally, for reversible filter banks we have $K=1$, in which case~(\ref{irr_normalization}) reduces to the requirement~(\ref{rev_normalization}).  

Note that, for reversible filter banks, the normalization~(\ref{lowpassDCconstraint}) represents the lowpass DC response of the unrounded, linear filter bank.  Moreover, this normalization must be implicit in the lifting filters themselves; it cannot simply be imposed by choosing the scaling factor $K=E_0(1)$.  Thus, there exist  dyadic filter banks that do \emph{not} satisfy the JPEG~2000 reversible normalization requirement~(\ref{rev_normalization}), e.g.,
\[
\mathbf{H}(z) =
\left[\begin{array}{cc}
1 & 1+z^{-1}\\
0 & 1
\end{array}\right]\,.
\]

\section{Uniqueness of Lifting Factorizations}\label{Uniqueness}
We now consider the relationship between lifting-domain gain normalizations and the apparently unrelated issue of uniqueness of filter banks.  Offhand, one doesn't expect any sort of uniqueness to hold for lifting factorizations since it is well-known that a single filter bank can be factored into many different lifting representations, e.g.
\begin{eqnarray}\label{identitylifting}
\mathbf{I}&=&
        \left[ \begin{array}{cc}
                1 & 0 \\
               -1 & 1
        \end{array}\right]
        \left[ \begin{array}{cc}
                1 & -1 \\
                0 & 1
        \end{array}\right]
        \left[ \begin{array}{cc}
                1       & 0 \\
                1/2    & 1
        \end{array}\right]
        \left[ \begin{array}{cc}
                1   & 2 \\
                0   & 1
        \end{array}\right]\\
&&\cdot
        \left[ \begin{array}{cc}
                1  & 0 \\
                -1/2  & 1
        \end{array}\right]
        \left[ \begin{array}{cc}
                1  & 1 \\
                0  & 1
        \end{array}\right]
        \left[ \begin{array}{cc}
                1  & 0 \\
                1  & 1
        \end{array}\right]
        \left[ \begin{array}{cc}
                1  & -1/2 \\
                0  & 1
        \end{array}\right]\,.\nonumber
\end{eqnarray}

\subsection{Linear Phase Group Lifting Factorizations}\label{Uniqueness:Linear}
\begin{figure}[t]
  \begin{center}
    \includegraphics{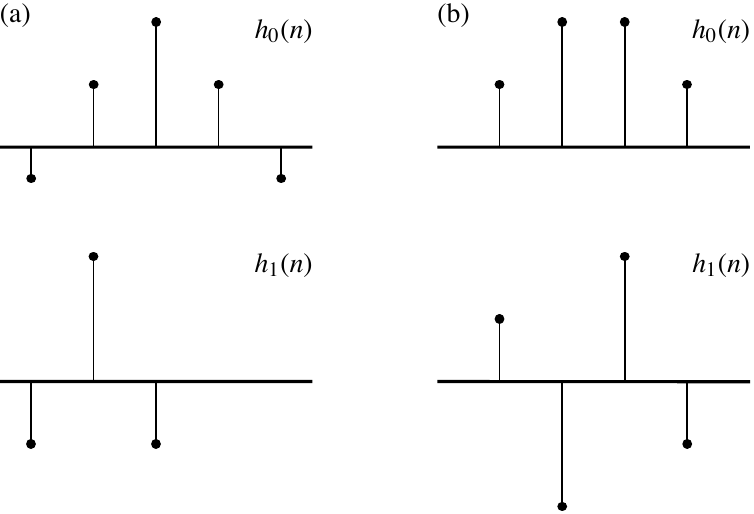}
    \caption{(a)~Whole-sample symmetric (WS) filter bank.\quad 
    (b)~Half-sample symmetric (HS) filter bank.}
    \label{WSHS}
  \end{center}
\end{figure}
For image coding, it is customary to use linear phase FIR filter banks.  Examples of the two principal classes of linear phase perfect reconstruction FIR filter banks, the whole-sample symmetric (WS) and half-sample symmetric (HS) classes, are indicated in Fig.~\ref{WSHS}.  It was shown in~\cite{BrisWohl06} that both classes always factor using linear phase lifting steps.  This analysis was based on the fact that linear phase lifting matrices form (multiplicative) matrix groups, and we refer to the resulting factorizations as \emph{group lifting factorizations}.     We summarize the results of~\cite{BrisWohl06} as follows.

{\bf Group Lifting Factorization of WS Filter Banks:} $\bfH(z)$ is WS if and only if it can be factored into a product of half-sample symmetric (FIR type 2) lifting steps,
\[ \bfH(z) = \mathbf{D}_K\,\bfS_{N-1}(z) \cdots \bfS_1(z)\,\bfS_0(z), \]
where $S_i(z)$ is symmetric about $1/2$ if $\bfS_i(z)$ is upper-triangular, or else  $S_i(z)$ is symmetric about $-1/2$ if $\bfS_i(z)$ is lower-triangular.

Note that~(\ref{identitylifting}) is \emph{not} a ``WS group lifting factorization'' of the WS filter bank $\mathbf{I}$ since it does not use HS lifting filters.  (Indeed, $\mathbf{I}=\mathbf{D}_1$ is already factored, trivially.)

The situation is more complicated for HS filter banks.

{\bf Group Lifting Factorization of HS Filter Banks:} $\bfH(z)$ is HS if and only if it can be lifted from a concentric equal-length HS ``base'' filter bank, $\bfB(z)$, by lifting filters  that are  whole-sample \emph{antisymmetric} about zero (FIR type 3, or WA),
\[ \bfH(z) = \mathbf{D}_K\,\bfS_{N-1}(z) \cdots \bfS_1(z)\,\bfS_0(z)\,\bfB(z). \]

The JPEG~2000 Part~2 standard specifies WS group lifting factorizations for WS filter banks in~\cite[Annex~G]{ISO_15444_2} and contains alternative specifications for non-WS filter banks in~\cite[Annex~H]{ISO_15444_2}.  (HS filter banks are not addressed specifically in Part~2, although they are allowed under the specifications of~\cite[Annex~H]{ISO_15444_2}, which contains examples of lifted HS filter banks that are effective for image coding applications.) 

While working on~\cite[Annex~G]{ISO_15444_2} the author observed, empirically, that there seemed to be one and \emph{only} one way to factor a given WS filter bank into a compliant WS group lifting factorization, which raised the question of whether the combination of the specialized group lifting framework and the Part~2 normalization requirements mathematically implies uniqueness of compliant WS group lifting factorizations.  The situation is far from simple, however, and we now outline the  theory presented in~\cite{Bris09,Bris09b}.

\subsection{Matrix Inner Automorphisms}
We begin with a simple computational formula arising from the group-theoretic perspective on lifting.  It is elementary to show that scaling matrices, $\bfD_K$, ``intertwine'' lifting matrices, by which we mean
\[ \mathbf{D}_K\,\mathbf{S}(z) = (\gam{K}\mathbf{S}(z))\,\mathbf{D}_K \,. \]
The matrix inner automorphism 
\[ \gam{K}\mathbf{A}(z) \equiv \mathbf{D}_K\,\mathbf{A}(z)\,\mathbf{D}_K^{-1} \]
has the action
\[  \gam{K}  \left[ \begin{array}{cc}
                    a  & b\\
                    c  & d
                    \end{array}\right]
\;=\; 
                    \left[ \begin{array}{cc}
                    a  & K^{-2} b \\
                    K^2 c  & d
                    \end{array}\right]\,.
\]

Start with the simplest possible example of a non-constant linear phase filter bank, the Haar filter bank, normalized according to~(\ref{lowpassDCconstraint}):
\[
\bfB_{haar}(z) \equiv
        \left[ \begin{array}{cc}
                1/2 & 1/2 \\
                -1 & 1
        \end{array}\right]\;=\;
        \left[ \begin{array}{cc}
                1  & 1/2 \\
                0  & 1
        \end{array}\right]
        \left[ \begin{array}{cc}
                1  & 0 \\
                -1  & 1
        \end{array}\right]\,.
\]
The Haar matrix has another factorization, 
\[
\bfB_{haar}(z) =
        \left[ \begin{array}{cc}
                1/2  & 0 \\
                0    & 2
        \end{array}\right]
        \left[ \begin{array}{cc}
                1  & 0 \\
                -1/2  & 1
        \end{array}\right]
        \left[ \begin{array}{cc}
                1  & 1 \\
                0  & 1
        \end{array}\right] \,,
        \]
which we write as 
\[\bfB_{haar}(z)=\bfD_2\,\bfB'_{haar}(z)\,, \]
where $\bfB'_{haar}(z)$ is a rescaled  Haar filter bank with $B'_0(1)=2$ and $B'_1(-1)=-1$.  

Let us show that this nonuniqueness propagates to any irreversible HS group lifting factorization lifted from the Haar ``base'' filter bank.  If
\[ \bfH(z) =\bfS_{N-1}(z) \cdots \bfS_1(z)\,\bfS_0(z)\,\bfB_{haar}(z) \]
is  lifted from the Haar using WA lifting filters, then 
\[H_0(1)=B_0(1)=1\]
by~(\ref{DC_gain_vector_recursion}) since $S_i(1)=0$ for all $i$, so $\bfH(z)$ is already normalized, regardless of whether the factorization is reversible or not.  

Replace $\bfB_{haar}(z)$ by $\bfD_2\,\bfB'_{haar}(z)$:
\begin{eqnarray*}
\bfH(z) &=& \bfS_{N-1}(z) \cdots \bfS_1(z)\,\bfS_0(z)\,\bfD_2\,\bfB'_{haar}(z)\\
&=& \bfD_2\,\bfS'_{N-1}(z) \cdots \bfS'_1(z)\,\bfS'_0(z)\,\bfB'_{haar}(z) 
\end{eqnarray*}
where 
\[\bfS_i(z)=\gam{2}\mathbf{S}'_i(z)\]
and $\bfB_{haar}(z)=\bfD_{2}\,\bfB'_{haar}(z)$.  Thus, at least in the irreversible case, we obtain two distinct  lifting factorizations of $\mathbf{H}(z)$.  We can readily generalize this example by defining
\[  \bfB^{(K)}_{haar}(z)\equiv \bfD_{1/K}\,\bfB_{haar}(z) \]
to obtain a continuum of such factorizations, each trivially different from the others.  

This merits a definition.

\textbf{Equivalence Modulo Rescaling: }  

Two lifting factorizations that differ only in this manner (i.e., via a matrix inner automorphism) are said to be \emph{equivalent modulo rescaling.}

Linear phase filter banks can still have linear phase lifting factorizations that are not equivalent modulo rescaling, however, such as the following example from~\cite{Bris09}.
\begin{eqnarray}
\mathbf{I}&=&\nonumber
        \left[ \begin{array}{cc}
                1   & (-{z^2}/4)(1-z^{-1})  \\
                0   & 1
        \end{array}\right]
        \left[ \begin{array}{cc}
                1       & 0\\
                -4(1+z^{-1})    & 1
        \end{array}\right]\\
        && \cdot\label{identity_lift2}
        \left[ \begin{array}{cc}
                1 & (5/4)(1-z^{-1}) \\
                0 & 1
        \end{array}\right]
        \left[ \begin{array}{cc}
                1 & 0\\
                -(1+z^{-1}) & 1
        \end{array}\right]\\
        && \cdot
        \left[ \begin{array}{cc}
                1 & -z^2(1-z^{-1})  \\
                0 & 1
        \end{array}\right]
        \left[ \begin{array}{cc}
                1 & 0\\
                5z^{-2}(1+z^{-1}) & 1
        \end{array}\right] \nonumber
\end{eqnarray}

\subsection{New Unique Factorization Results}
Given such examples, it is rather surprising that one can nonetheless prove unique lifting factorization results for both WS and HS \emph{group lifting factorizations}.  Note that~(\ref{identity_lift2}) is \emph{not} a WS group lifting factorization in the sense defined above since it involves both half-sample symmetric \emph{and} half-sample  antisymmetric lifting filters.  This breaks the  group-theoretic underpinnings of the following uniqueness results, which can be found in~\cite{Bris09b}.  The proofs are somewhat nontrivial and involve the unfortunately complicated notion of  \emph{group lifting structures} so we will not attempt to describe them here.  Instead we summarize the main results of~\cite{Bris09b} using the terminology already defined.

\textbf{Uniqueness Theorems for Linear Phase Group Lifting Factorizations: } 

(1) Any two irreversible HS group lifting factorizations of the same HS filter bank are equivalent modulo rescaling.  

(2) Any two \emph{reversible} HS group lifting factorizations are identical.  

(3) Any two WS group lifting factorizations of the same WS filter bank (reversible or irreversible) are identical.  

\rem  The third result implies that any two lifting factorizations of WS filter banks that comply with JPEG~2000 Part~2 Annex~G syntax are necessarily the same, confirming the author's observations mentioned above.  No such results hold for Annex~H syntax, which does not restrict the structure of lifting factorizations in any way.  Thus, although HS filter banks can be specified using Annex~H syntax, there are no group-theoretic constraints to preclude the use of other liftings besides HS group lifting factorizations.

The \emph{failure} of the group lifting factorization  approach to yield uniqueness results  for paraunitary filter banks, in spite of their natural group-theoretic structure, is discussed in~\cite{Bris09}.  The basic problem is that the algebraic structure of lifting factorizations for paraunitary filter banks does not structurally enforce the paraunitary property of the filter banks.

\newpage



\begin{thebibliography}{1}
\providecommand{\url}[1]{#1}
\csname url@samestyle\endcsname
\providecommand{\newblock}{\relax}
\providecommand{\bibinfo}[2]{#2}
\providecommand{\BIBentrySTDinterwordspacing}{\spaceskip=0pt\relax}
\providecommand{\BIBentryALTinterwordstretchfactor}{4}
\providecommand{\BIBentryALTinterwordspacing}{\spaceskip=\fontdimen2\font plus
\BIBentryALTinterwordstretchfactor\fontdimen3\font minus
  \fontdimen4\font\relax}
\providecommand{\BIBforeignlanguage}[2]{{%
\expandafter\ifx\csname l@#1\endcsname\relax
\typeout{** WARNING: IEEEtran.bst: No hyphenation pattern has been}%
\typeout{** loaded for the language `#1'. Using the pattern for}%
\typeout{** the default language instead.}%
\else
\language=\csname l@#1\endcsname
\fi
#2}}
\providecommand{\BIBdecl}{\relax}
\BIBdecl

\bibitem{DaubSwel98}
I.~C. Daubechies and W.~Sweldens, ``Factoring wavelet transforms into lifting
  steps,'' \emph{J.~Fourier Anal.\ Appl.}, vol.~4, no.~3, pp. 245--267, 1998.

\bibitem{BrisWohl06}
C.~M. Brislawn and B.~Wohlberg, ``The polyphase-with-advance representation and
  linear phase lifting factorizations,'' \emph{IEEE Trans.\ Signal Process.},
  vol.~54, no.~6, pp. 2022--2034, Jun. 2006.

\bibitem{Sweldens96}
W.~Sweldens, ``The lifting scheme: a custom-design construction of biorthogonal
  wavelets,'' \emph{Appl.\ Comput.\ Harmonic Anal.}, vol.~3, no.~2, pp.
  186--200, 1996.

\bibitem{ISO_15444_1}
\emph{Information technology---{JPEG}~2000 {I}mage {C}oding {S}ystem,
  {P}art~1}, ser. ISO/IEC Int'l.\ Standard~15444-1, ITU-T Rec.~T.800.\hskip 1em
  plus 0.5em minus 0.4em\relax Int'l.~Org.\ for Standardization, Dec. 2000.

\bibitem{TaubMarc02}
D.~S. Taubman and M.~W. Marcellin, \emph{JPEG2000: Image Compression
  Fundamentals, Standards, and Practice}.\hskip 1em plus 0.5em minus
  0.4em\relax Boston, MA: Kluwer, 2002.

\bibitem{ISO_15444_2}
\emph{Information technology---{JPEG}~2000 {I}mage {C}oding {S}ystem, {P}art~2
  ({E}xtensions)}, ser. ISO/IEC Int'l.\ Standard~15444-2, ITU-T
  Rec.~T.801.\hskip 1em plus 0.5em minus 0.4em\relax Int'l.~Org.\ for
  Standardization, May 2004.

\bibitem{BrisWohl07}
C.~M. Brislawn and B.~Wohlberg, ``Gain normalization of lifted filter banks,''
  \emph{Signal Processing}, vol.~87, no.~6, pp. 1281--1287, Jun. 2007.

\bibitem{Bris09}
C.~M. Brislawn, ``Group lifting structures for multirate filter banks,~{I}:
  {U}niqueness of lifting factorizations,'' Los Alamos National Lab, Tech. Rep.
  LAUR-09-2983, May 2009, submitted for publication.

\bibitem{Bris09b}
------, ``Group lifting structures for multirate filter banks,~{II}: {L}inear
  phase filter banks,'' Los Alamos National Lab, Tech. Rep. LAUR-09-3049, May
  2009, submitted for publication.

\end{thebibliography}
\end{document}